\documentclass[runningheads]{llncs}
\usepackage[T1]{fontenc}
\usepackage{graphicx}

\usepackage{graphicx}
\usepackage{array}
\usepackage{booktabs}
\usepackage{multirow}
\usepackage{float}

\usepackage{pgfplots}
\usepackage{pgfplotstable}
\usepackage{tikz}

\begin{document}
\title{Evaluating Large Language Models for the Generation of Unit Tests with Equivalence Partitions and Boundary Values}
\titlerunning{Evaluating Large Language Models for the Generation of Unit Tests}
\author{Martín Rodríguez\inst{1}\orcidID{0009-0001-2755-3575} \and
Gustavo Rossi\inst{1}\orcidID{0000-0002-3348-2144} \and
Alejandro Fernandez\inst{1}\orcidID{0000-0002-7968-6871}}
\authorrunning{Martín Rodríguez et al.}

\institute{LIFIA, Faculty of Informatics, UNLP, La Plata, Argentina}
\maketitle              
\begin{abstract}
The design and implementation of unit tests is a complex task many programmers neglect. This research evaluates the potential of Large Language Models (LLMs) in automatically generating test cases, comparing them with manual tests. An optimized prompt was developed, that integrates code and requirements, covering critical cases such as equivalence partitions and boundary values. The strengths and weaknesses of LLMs versus trained programmers were compared through quantitative metrics and manual qualitative analysis. The results show that the effectiveness of LLMs depends on well-designed prompts, robust implementation, and precise requirements. Although flexible and promising, LLMs still require human supervision. This work highlights the importance of manual qualitative analysis as an essential complement to automation in unit test evaluation.
\end{abstract}

\section{Introduction}

Software testing is a fundamental stage in software development. It allows for the identification and correction of errors before the product reaches production, ensuring its quality and expected performance. Test automation optimizes this process by reducing the time and resources, providing consistent and repeatable results \cite{gomez2021tests}. Within this approach, unit testing plays a key role by evaluating small code units in isolation, facilitating early bug fixes and ensuring that each component works correctly before being integrated into the complete system \cite{Koomen1999}.

Writing unit tests and identifying relevant test cases is tedious. This leads many programmers to skip this step altogether. Studies have shown that numerous projects on GitHub lack test files \cite{gonzalez2017}. Missing unit tests compromise software quality and overload the QA team, limiting their ability to perform other essential checks. Faced with this challenge, there has been a growing interest in automated code generation and testing using Large Language Models (LLMs) to optimize the process and reduce the manual workload.

The use of AI tools in software development has increased significantly. In 2024 Stack Overflow Developer's Survey \cite{StackOverflow2024AI} of 60,907 developers, 76\% said they use or plan to use AI in their workflows by 2025. Of the 34,168 respondents about its impact in different areas, 81\% expect more integration in code documentation, 80\% in testing, and 76\% in writing code. This trend is also reflected in interviews with entrepreneurs who manage development teams, who note that developers are looking to delegate repetitive tasks to AI, such as test creation, and then focus on more creative activities 
\cite{stackoverflow2024genai}. In addition, they point out that generating more lines of code does not guarantee better quality; producing clear, efficient code with fewer errors is preferable.

LLMs have the potential to generate unit tests faster, improving code coverage, identifying hard-to-detect cases, and dynamically adapting to changes in the source code. While a developer designs, writes, and verifies each test manually, these models could produce multiple test cases in seconds. This capability has been demonstrated in large-scale content generation in other contexts \cite{tamkin2021understandingcapabilitieslimitationssocietal}, suggesting that it could also be applied to unit test automation. As studies on LLMs advance in the field, the question of how they compare with human experts remains open. This article discusses LLM's accuracy, consistency, and limitations in test generation, focusing on two specific test case selection strategies: boundary values, equivalence partitions. 

Traditionally, software engineers design and write unit tests manually from an analysis of the functional and non-functional requirements of the system, ensuring that the software meets the user's specifications and needs. They create detailed test cases with input conditions, execution steps, output expectations and success or failure criteria. There are several strategies for identifying effective test cases \cite{Basili2006}. In this work, two of them are of interest: equivalence partitions and boundary values. An equivalence partition divides the input data into classes with similar behavior, allowing the number of tests to be reduced by selecting one representative value per partition. If one case fails, it is assumed that the others in the same class will fail as well \cite{graham2008foundations}. Boundary value analysis complements this technique, as errors are often concentrated at the boundaries of the input domain \cite{pressman2005ingenieria}. Maximum and minimum values, including both valid and invalid values, are tested to detect faults in boundary conditions \cite{graham2008foundations}. These strategies not only optimize test case selection, but also prioritize test quality and effectiveness over simply increasing code coverage. Despite their simplicity, these strategies are both effective and widely adopted, and often challenging enough to justify the use of AI assistance in practice. 

This article's contributions are:
\begin{enumerate}
    \item A set of software artifacts for varied scenarios in unit testing.
    \item Benchmark tests created by expert programmers, based on boundary values and equivalence partitions.
    \item LLM prompts for automated test generation (based on a common template).
    \item Comparative quantitative and qualitative evaluation of LLM-generated tests against manually developed tests.
    \item A critical discussion of the strengths and limitations of LLMs in test generation.
\end{enumerate}

Contributions 1 to 3 are publicly available as a Git repository \cite{datasetSJL5HU_2025}, at the National University of La Plata data repository. 

The paper is organized as follows: Section \ref{section:related-work} provides an overview of LLMs and their application in code generation, as well as a comparison with traditional approaches used by software engineers to write unit tests. In addition, previous work in automatic test generation is reviewed, and the gap this study seeks to address is identified. Section \ref{section:experimental-design} describes the experiment setup, tools, participants, and evaluation metrics used to compare human-generated tests and LLMs. Section \ref{section:results} focuses on the quantitative evaluation, while Section \ref{section:discussion} focuses on the qualitative evaluation. Section \ref{section:threats} discusses possible limitations of the study. Finally, Section  \ref{section:conclusions} summarizes the main contributions of the article and proposes directions for future research in automated test generation using LLMs.

\section{Related Work}
\label{section:related-work}

Guilherme et al. \cite{guillerme2023} used GPT-3.5-turbo to generate unit tests on 33 programs in a fully automated way. In their study, they design a prompt to request tests for an entire class, experimenting with different levels of `temperature'. The prompt they use only includes indications on the format and requirements of the tests, superficially mentioning the use of boundary values, without considering it in the final evaluation. In this research, the application of equivalence partitions and boundary values is addressed more rigorously, in addition to evaluating the results with a catalog of criteria that measures the efficiency in the coverage of the cases.

Although Guilhermes et al.'s results show good performance in code coverage, mutation score, and test success rate, they do not include the requirements specification in the prompt, which may be key to the LLM's accuracy. This paper considers it essential to include such specification so that the model better interprets the context and does not rely on code alone, which may contain errors from the outset. Unlike the Guilhermes et al. approach, which does not compare automatically generated tests with manual tests, this article considers such a comparison essential. Tests created by programmers can capture specific nuances of the problem domain that an automated tool may miss.

Max Schafer et al.\cite{schäfer2023} address some limitations of the work of Guilhermes et al. with TestPilot, a tool that generates unit tests using commercial LLMs. In this case, tests are requested for a specific function. In addition to indicating the test format, the prompts include detailed information about the function, including its signature, documentation comments, usage examples, source code, and test scaffolding. If a test fails, a new request is made incorporating the error message and the failed test in search of a fix. This approach evaluates the coverage of statements and branches, as well as the success rate and reasons for failure. While this work overcomes some limitations of Guilherme's approach, it still does not apply equivalence partitions and boundary values, nor does it compare automatically generated tests with manual tests.

There are cases where the perspective of a human developer is taken into account. Siddiq et al.\cite{Siddiq_2024} investigate the effectiveness of LLMs in generating unit tests for software, exploring their ability to understand functional requirements and generate relevant test cases. The LLM is provided with a context with the class code including imports, along with a prompt specifying the name of the test file, the method under test (MUT), and details on the number and types of tests to be generated. The study analyses the quality, diversity, and coverage of the tests generated from the humanEval data corpus \cite{DuXueying24}. Unlike previous work, it compares these tests with those generated by technologies such as EvoSuite \cite{evoSuite2011} and manual tests. However, as in previous work, it does not consider the criterion of equivalence partitions or boundary values for the tests.

Most studies focus on maximizing code coverage. However, Guilherme and his team are the only ones to mention the inclusion of boundary values, albeit superficially and without concrete analysis. In contrast, Schafer and Siddiq are notable for including the program specification in the prompt, which allows for a more detailed and grounded approach.

Siddiq and his team are the only ones who, like this work, compare tests generated by LLMs with manual tests performed by humans, while other studies evaluate the effectiveness of the tests in different ways. Guilhermes and his team compare results with reference sets using EvoSuite and other traditional tools, using PiTest \cite{PIT2016} for mutation score and code coverage reports. On the other hand, Schafer and his team evaluate the coverage achieved by their tests against Nessie\cite{Nessie2022}, using tools such as lstanbul/nyc (https://istanbul.js.org/) to measure statement and branch coverage, and checking for the absence of trivial assertions with CodeQL\cite{CodeQL_2023}.

This work seeks to combine approaches to obtain more complete and meaningful results. It relies on the capability of commercially available general-purpose LLMs to create unit tests based on equivalence partitions and boundary values, developing a prompt that includes all the necessary information, such as the requirements specification and the code under test. In addition, the results obtained are compared with manual tests generated by a trained developer to validate the efficiency and quality of the automatically generated tests.

\section{Experimental Design}
\label{section:experimental-design}

\begin{figure}
    \centering
    \includegraphics[width=\linewidth]{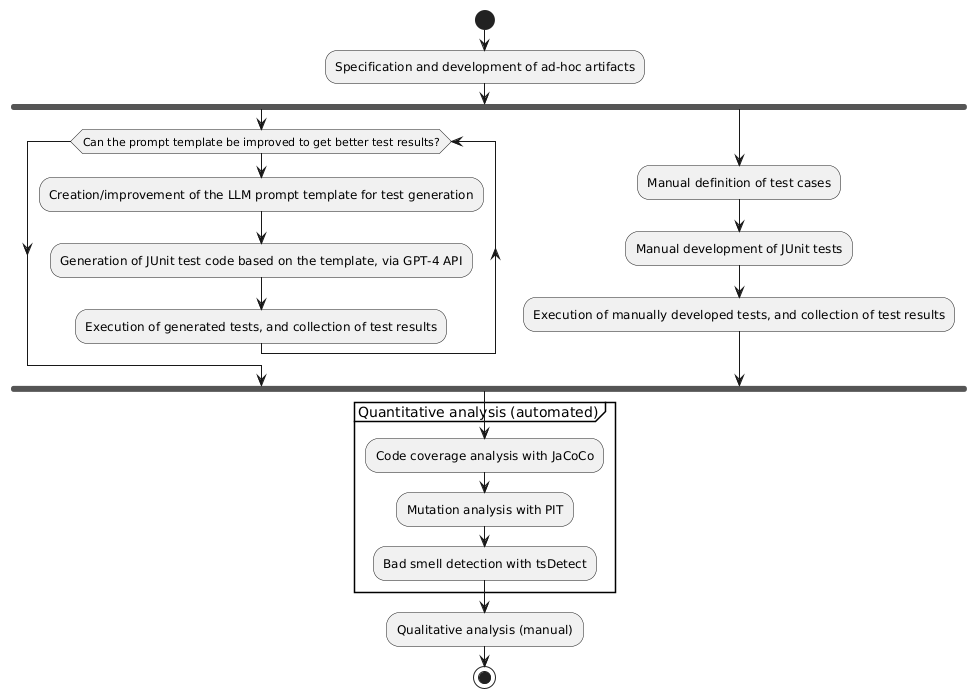}
    \caption{Overview of the experiment's design}
    \label{fig:experiment-design-diagram}
\end{figure}

This paper compares the quality of tests generated automatically by an LLM with those elaborated manually by developers for an ad hoc selection of software artifacts. Figure \ref{fig:experiment-design-diagram} presents the general design of the experiment.

The reference technology is the Java language and the JUnit framework for unit test automation. JUnit is one of the most widely used tools for unit test automation in Java, and has been fundamental in adopting methodologies such as test-driven development (TDD). 

OpenAI's GPT-4-0613 model is chosen as an LLM case study because it handles natural language and complex concepts, facilitating the generation of tests aligned with the requirements of each artifact. In addition, its API allows for flexible integration and quick adjustments to prompts, enabling the systematization of the experiment.

A representative set of Java artifacts is needed as a basis for testing for the study of equivalence partitions and boundary values. While the possibility of using software artifacts that had been the focus of study in previous literature was evaluated, the literature analysis did not result in suitable candidates. Consequently, it was decided to develop an ad hoc set of test artifacts.

The design of prompts is a critical stage in this type of experiment. A template for generating prompts is iteratively developed until the prompts result in effective, efficient, and robust tests (on the left in Figure \ref{fig:experiment-design-diagram}). The template takes as input the specification and implementation of an artifact, and returns a prompt that asks the LLM to generate the test. A template was used (instead of one prompt per artifact) to aim at generalizable results. In parallel, programmers with knowledge in test case design write tests for the same artifacts following criteria of equivalence partitions and boundary values (on the right in Figure \ref{fig:experiment-design-diagram}). 

Replicating the evaluation methods used in similar research, the tests are compared regarding code coverage, mutation score, and bad smell detection. In addition, and unlike in other works, a manual qualitative analysis is performed to evaluate effectiveness in detecting equivalence partitions, boundary values, exception handling, and more subjective aspects such as code clarity. 

\subsection{Artifacts to be tested}

In object-oriented programming, the method is the minimum unit to be tested. The test case search space for a method is defined by the parameters of the method and by the internal state of the object (including all possible combinations of these). The post-conditions of a test (controlled in the asserts) are given by the final state of the object after the test, the value returned by the method, and the occurrence or non-occurrence of exceptions. Whoever defines test cases must identify equivalence partitions and boundary values in that universe.

For the present experiment, a set of software artifacts to be tested must capture representative, non-redundant, complete, yet simple test situations. In this work, we recognize five scenarios to evaluate the ability to identify and develop test cases for an LLM and a developer. 

\begin{enumerate}
    \item Variations on a single parameter: Evaluates how the LLM detects equivalence partitions and boundary values for a single input. 
    \item Correlated variations of two parameters: Introduces complexity by requiring the handling of interactions between both parameters, demanding precision in the creation of partitions and selection of boundary values.  
    \item Variations on a single instance variable: Introduces object state into the equation, assessing the LLM's ability to correctly define and instantiate objects and consider the impact of internal state on results.  
    \item Correlated variations of a parameter and an instance variable: Combines external input and internal state, increasing complexity. Evaluates whether the LLM handles this interaction properly and selects the correct boundary values for both (and/or the correlation).  
    \item Correlated variations of two instance variables: This is the most complex scenario, where both variables are part of the internal state. It checks if the LLM correctly handles its interaction over time and covers all combinations of boundary values.
\end{enumerate}

Additionally, for each scenario, two variants are identified: one where the program chooses between two alternative execution paths and one where the alternatives are to execute one path or throw an exception. This allows assessing whether the LLM covers all execution paths and handles exceptions correctly.  

An ad hoc set of software artifacts was developed covering the 10 scenarios described above. These modular and minimal artifacts contain methods that handle inputs, outputs, and exceptions, with precise functional purposes, such as checking parity, validating passwords, operating bank accounts, or manipulating collections. In addition to generating outputs, they also throw exceptions under certain conditions.

The source code for the set of scenarios was organized in a Git repository. Each scenario, from scenario01 to scenario10, is located in a clearly and consistently named folder. Within each, a Markdown file with its respective functional specification is included. 

\subsection{Iterative development of prompts}

The approach used in this work aims to make prompts clear and accurate and minimize errors in the generated tests. Prompts are engineered with specific instructions, iteratively refining their formulation to improve accuracy. The quality of test suites is assessed by code coverage, mutation scoring, bad smells detection, and qualitative (manual) analysis.

Role-prompting was used as a strategy for prompt design, assigning the LLM the role of a Java developer with 25 years of testing experience with JUnit, equivalence partitions, and boundary values. OpenAI's prompting recommendations were followed, such as including context, task, examples, and \cite{openai2024promptengineering} formatting. The context indicates that the requester has already implemented the classes, but does not have time to write the tests. The task is to generate JUnit tests by applying equivalence partitions and boundary values with clear documentation. In addition, an example of identifying these values and the structure the tests should follow is provided.

The prompt includes the specification and implementation of the artifact to be tested to ensure adequate testing. Although the implementation is not essential in early stages (as in TDD), it provides context for the LLM to generate more complete tests. The specification defines what is to be tested, ensuring that the tests validate the system's expected behavior.

The Git repository includes a base template (promptTemplate.md) that is manually adjusted in each iteration, and a Python script that automates the transformation of the template into a prompt for each artifact (implementation and specification). With the template transformed into a specific prompt, the script sends the prompts to the GPT-4 API and stores the generated tests in the corresponding folder under the format ClassNameTest\_GPT.java.

One of the main challenges is getting the model to correctly identify equivalence partitions and boundary values, avoiding irrelevant tests. Although in some cases it detects them correctly, in others, the tendency to select intermediate values persists despite multiple iterations of improvement of the prompts template. After multiple iterations of improvement, the version included in the repository (and against which no substantial improvements could be obtained) was reached. 

Finally, after multiple script runs with the optimized prompt template, three are randomly selected to ensure an unbiased sample (in the rest of the article, we refer to the selected runs as Run 1, 2, and 3, respectively). The final set consists of 40 test classes: 10 manual and 30 automatically generated (three per scenario). The repository is organized in four branches: main, which contains the manual tests, and branches branch01, branch02, and branch03, where the run results are recorded. Each API query is performed in a separate context to avoid bias between scenarios. In addition, a temperature of 0.3 is used to ensure accuracy and reduce unexpected variations in the generated code.

\subsection{Selection and instruction of participants}

For the manual development of tests (against which the tests generated by the LLM will be compared), a developer with expertise in identifying equivalence partitions and boundary values is invited. The developer is given specific instructions to avoid bias and standardize the tests, prioritizing quality over quantity. He/she is asked to apply techniques of equivalence partitions and boundary values, use JUnit, maintain consistent nomenclature in the naming of classes (classWhichIsTesting\_Human.java), and adequately document each test. Instead of maximizing code coverage, the developer is instructed to focus on designing representative cases of system behavior. The developer has access to the Git repository.

\subsection{Metrics and evaluation}

To compare automatically generated test cases with manually designed ones, we start with quantitative metrics that allow an objective assessment of their quality. In particular, coverage analysis, mutation testing, and automatic detection of bad smells are combined. The combination of coverage analysis and mutation testing allows not only the detection of untested code but also the evaluation of the effectiveness of the tests in identifying bugs. Test smells are design problems in tests that can compromise their reliability. These include instability, false positives or negatives, code duplication, and lack of documentation in assertions, which affect the ability to detect defects.

For coverage analysis, we used the JaCoCo (https://www.eclemma.org/jacoco/) tool, which measures the percentage of executed code, including branch coverage, to identify untested and potentially bug-prone areas. For mutation testing, we used the PIT \cite{PIT2016} tool which introduces deliberate changes to the code (mutations) to assess the robustness of the tests: if they fail to detect the mutation, it survives, indicating a possible weakness; if they identify it and fail, the mutation is considered to be eliminated. TsDetect \cite{tsdetect2020} was used to detect test smells.

The quantitative analysis is complemented by a \textbf{qualitative analysis} conducted manually by the authors, due to its subjective nature. This analysis seeks to address the inherent limitations of automated metrics by providing a more in-depth assessment of test performance. Incorporating human judgment allows for identifying subtle issues of readability, clarity, and consistency in test nomenclature and documentation.

The comparison criteria for the manual analysis are as follows:
\begin{itemize}
\item Equivalence partitions: assess whether both approaches correctly identify and cover the input partitions.
\item Boundary values: check whether the values at the upper and lower boundaries of each partition are tested.
\item Exception handling: analyze whether exceptions for out-of-bounds or invalid values are properly considered.
\item Numerical inaccuracies: assess whether precision errors in data types such as double or float (e.g., use of delta in assertEquals) are handled correctly.
\item Code structure: examine the clarity and consistency in the organization of the test code.
\item Consistent and descriptive names: ensure that test classes reflect the tested functionality and that method names clearly express the tested behavior.
\item Documentation: review the inclusion of comments and documentation that facilitate the understanding of the tests.
\end{itemize}

\section{Quantitative evaluation}
\label{section:results}

Analyzing the results of the coverage tests (Table \ref{tab:coverage_results}), both the GPT-generated and manual tests achieve 100\% branch coverage in all scenarios evaluated. This indicates that all equivalence partitions were identified and executed correctly. However, scenario 06 in Run 1 has a coverage of 0.0\% due to an error in invoking the non-existent `setValue' method, which prevents the tests from running.  

In terms of line coverage, some scenarios do not reach 100\%, such as scenario 06 of Run 1, where only configuration and `setUp' lines are executed, without covering functional code. However, in other cases, lower coverage is not negative, but results from the exclusion of classes such as context, which is not essential for the evaluated logic. For example, in scenario 02, where the main logic resides in the `passwordStrategy' strategy, the tests focus on this, leaving out the context.  

It is observed that, in scenarios 01 to 04, GPT includes context unnecessarily, leading to 100\% coverage due to an isolation error. In contrast, the manual tests and scenario 04 of Runs 1 and 3 avoid testing the context and focus on the strategy, which reduces their line coverage. This does not affect branch coverage, as the context does not contain forks.

\begin{table}
    \centering
    \renewcommand{\arraystretch}{1.2} 
    \small 
    \resizebox{\textwidth}{!}{ 
    \begin{tabular}{|c|c|c|c|c|c|c|c|c|}
        \hline
        \multirow{2}{*}{\textbf{Scenario}} & \multicolumn{2}{c|}{\textbf{Run 1}} & \multicolumn{2}{c|}{\textbf{Run 2}} & \multicolumn{2}{c|}{\textbf{Run 3}} & \multicolumn{2}{c|}{\textbf{Manual}} \\
        \cline{2-9}
        & \textbf{Branch Cov.} & \textbf{Line Cov.} & \textbf{Branch Cov.} & \textbf{Line Cov.} & \textbf{Branch Cov.} & \textbf{Line Cov.} & \textbf{Branch Cov.} & \textbf{Line Cov.} \\
        \hline
        \textbf{01} & 100\% & 100\% & 100\% & 100\% & 100\% & 100\% & 100\% & 50\% \\
        \textbf{02} & 100\% & 100\% & 100\% & 100\% & 100\% & 100\% & 100\% & 33.3\% \\
        \textbf{03} & 100\% & 100\% & 100\% & 100\% & 100\% & 100\% & 100\% & 73.3\% \\
        \textbf{04} & 100\% & 69.2\% & 100\% & 100\% & 100\% & 69.2\% & 100\% & 69.2\% \\
        \textbf{05} & 100\% & 100\% & 100\% & 100\% & 100\% & 100\% & 100\% & 100\% \\
        \textbf{06} & 0.0\% & 75\% & 100\% & 100\% & 100\% & 100\% & 100\% & 100\% \\
        \textbf{07} & 100\% & 100\% & 100\% & 100\% & 100\% & 100\% & 100\% & 100\% \\
        \textbf{08} & 100\% & 100\% & 100\% & 100\% & 100\% & 100\% & 100\% & 100\% \\
        \textbf{09} & 100\% & 100\% & 100\% & 100\% & 100\% & 100\% & 100\% & 100\% \\
        \textbf{10} & 100\% & 100\% & 100\% & 100\% & 100\% & 100\% & 100\% & 100\% \\
        \hline
    \end{tabular}
    } 
    \caption{Branch and line coverage analysis for each run and scenario.}
    \label{tab:coverage_results}
\end{table}

Analysis of the mutation test results in Table \ref{tab:mutation_results} shows that, in most scenarios, both GPT-generated and manual tests achieve 100\% mutation coverage, indicating that all possible mutations in the code are detected and handled appropriately.  

However, some runs do not generate mutations because PIT only executes them if all tests are completed without errors. In this context, four GPT tests have problems: one in scenario 06 of Run 1, already mentioned, and three more due to errors in floating point value comparisons.  

Furthermore, in scenarios 01 to 04, the manual tests show lower percentages due to the lack of coverage in the context classes, as observed in the JaCoCo results. Despite this, the 100\% coverage in other cases indicates that each mutation is tested at least once, suggesting that the tests are effective and reliable in detecting unintended changes in the code.

\begin{table}
    \centering
    \renewcommand{\arraystretch}{1.2} 
    \small 
    \resizebox{\textwidth}{!}{ 
    \begin{tabular}{|c|c|c|c|c|c|c|c|c|}
        \hline
        \multirow{2}{*}{\textbf{Scenario}} & \multicolumn{2}{c|}{\textbf{Run 1}} & \multicolumn{2}{c|}{\textbf{Run 2}} & \multicolumn{2}{c|}{\textbf{Run 3}} & \multicolumn{2}{c|}{\textbf{Manual}} \\
        \cline{2-9}
        & \textbf{Line Cov.} & \textbf{Mut. Cov.} & \textbf{Line Cov.} & \textbf{Mut. Cov.} & \textbf{Line Cov.} & \textbf{Mut. Cov.} & \textbf{Line Cov.} & \textbf{Mut. Cov.} \\
        \hline
        \textbf{01} & 100\% & 100\% & 100\% & 100\% & 100\% & 100\% & 50\% & 63\% \\
        \textbf{02} & 100\% & 100\% & 100\% & 100\% & 100\% & 100\% & 33\% & 63\% \\
        \textbf{03} & 100\% & 100\% & 100\% & 100\% & 100\% & 100\% & 73\% & 88\% \\
        \textbf{04} & 69\% & 89\% & 100\% & 100\% & 69\% & 78\% & 69\% & 88\% \\
        \textbf{05} & 100\% & 100\% & 100\% & 100\% & 100\% & 100\% & 100\% & 100\% \\
        \textbf{06} & - & - & 100\% & 100\% & 100\% & 100\% & 100\% & 100\% \\
        \textbf{07} & 100\% & 100\% & 100\% & 100\% & 100\% & 100\% & 100\% & 100\% \\
        \textbf{08} & 100\% & 100\% & - & - & - & - & 100\% & 100\% \\
        \textbf{09} & 100\% & 100\% & 100\% & 100\% & 100\% & 100\% & 100\% & 100\% \\
        \textbf{10} & - & - & 100\% & 100\% & 100\% & 100\% & 100\% & 100\% \\
        \hline
    \end{tabular}
    } 
    \caption{Line and mutation coverage for each run and scenario.}
    \label{tab:mutation_results}
\end{table}

The test smells analysis identifies bad smells in the test code, examining 21 types, of which four were detected:  Assertion Roulette (a test method contains multiple assertions without documentation, affecting readability and making it difficult to identify bugs); General Fixture (a test fixture is too general and only partially used in test methods); Magic Number Test (numeric literals are used in assertions without context, making it difficult to understand and maintain code); Duplicate Assert (a test checks the same condition multiple times in a single method, generating redundancy).

The results are presented in Table \ref{tab:resultados_categorias_smells}. Assertion Roulette was detected in GPT-generated and manually generated tests, suggesting a recurring problem.  General Fixture was reported in two manual tests, where the `setUp' method instantiated unused fields in all tests.  Magic Number Test was found in all scenarios tested by GPT and manually, indicating a common problem.  Duplicate Assert was identified in two GPT runs for the same scenario, suggesting a trend in automatic test generation. 

\begin{table}
    \centering
    \setlength{\tabcolsep}{5pt} 
    \renewcommand{\arraystretch}{1.2} 
    \small 
    \resizebox{\textwidth}{!}{ 
        \begin{tabular}{|c|c|c|c|c|c|c|c|c|c|c|c|c|c|c|c|c|}
            \hline
            \multirow{2}{*}{\textbf{Scenario}} & \multicolumn{4}{c|}{\textbf{Assertion Roulette}} & \multicolumn{4}{c|}{\textbf{General Fixture}} & \multicolumn{4}{c|}{\textbf{Magic Number Test}} & \multicolumn{4}{c|}{\textbf{Duplicate Assert}} \\
            \cline{2-17}
            & \textbf{B1} & \textbf{B2} & \textbf{B3} & \textbf{Man} & \textbf{B1} & \textbf{B2} & \textbf{B3} & \textbf{Man} & \textbf{B1} & \textbf{B2} & \textbf{B3} & \textbf{Man} & \textbf{B1} & \textbf{B2} & \textbf{B3} & \textbf{Man} \\
            \hline
            \textbf{01} & 2 & 0 & 2 & 1 & 0 & 0 & 0 & 0 & 4 & 4 & 2 & 3 & 0 & 0 & 0 & 0 \\
            \textbf{02} & 0 & 0 & 2 & 0 & 0 & 0 & 0 & 0 & 3 & 3 & 3 & 2 & 0 & 0 & 0 & 0 \\
            \textbf{03} & 0 & 0 & 0 & 2 & 0 & 0 & 0 & 3 & 3 & 3 & 3 & 3 & 0 & 0 & 0 & 0 \\
            \textbf{04} & 0 & 0 & 0 & 0 & 0 & 0 & 0 & 3 & 3 & 3 & 3 & 3 & 0 & 0 & 0 & 0 \\
            \textbf{05} & 3 & 2 & 1 & 2 & 0 & 0 & 0 & 0 & 4 & 4 & 4 & 4 & 1 & 0 & 1 & 0 \\
            \textbf{06} & 0 & 0 & 0 & 0 & 0 & 0 & 0 & 0 & 5 & 5 & 5 & 4 & 0 & 0 & 0 & 0 \\
            \textbf{07} & 3 & 3 & 4 & 2 & 0 & 0 & 0 & 0 & 3 & 3 & 4 & 2 & 0 & 0 & 0 & 0 \\
            \textbf{08} & 4 & 4 & 5 & 2 & 0 & 0 & 0 & 0 & 4 & 4 & 5 & 2 & 0 & 0 & 0 & 0 \\
            \textbf{09} & 3 & 2 & 4 & 2 & 0 & 0 & 0 & 0 & 3 & 4 & 4 & 2 & 0 & 0 & 0 & 0 \\
            \textbf{10} & 5 & 3 & 3 & 2 & 0 & 0 & 0 & 0 & 5 & 3 & 3 & 2 & 0 & 0 & 0 & 0 \\
            \hline
        \end{tabular}
    }
    \caption{Comparison of bad smells between automated runs and manual analysis.}
    \label{tab:resultados_categorias_smells}
\end{table}

\section{Qualitative evaluation}
\label{section:discussion}  

Both approaches (automatic and manual) correctly identify equivalence partitions. However, a manual analysis reveals differences in the handling of boundary values. In general, manual tests tend to be more exhaustive, using values as close as possible to the edges of each partition, which allows for a more accurate verification of the system. On the other hand, GPT-generated tests tend to use representative values within partitions, usually towards the middle of the range, which provides acceptable coverage but does not accurately capture immediate edge cases. This difference is relevant, as values close to the edge can generate different behaviors than central values, highlighting the importance of adequate coverage.

In some cases where GPT tests boundary values, failures are detected in comparisons with floating-point numbers due to slight decimal differences. These errors arise from the lack of a proper tolerance (delta) in the assertions, an essential practice to avoid inaccuracies in calculations with floats. In contrast, manual tests do not face problems with floating-point values, as they avoid direct comparisons in this format, unlike those generated by GPT. This reveals a difference in approach: the manual tests are more narrow and precise, while the GPT tests cover a wider range. For example, GPT tests values below, at, and above the edge, while the manual tests focus only on values at and above the edge. Thus, manual tests prioritize accuracy over quantity, avoiding unnecessary cases and improving efficiency.

In scenarios 01 to 04, the GPT-generated tests do not always directly test the class implementing the logic, but instantiate the context class with the strategy (these scenarios are implementations of the Strategy design pattern). Although this ensures adequate coverage and evaluates equivalence partitions and boundary values, it does not follow the good practice of testing each class separately.  The context class orchestrates strategies, while the main logic resides in the strategy classes, so unit tests must focus on each in isolation. GPT achieves this in some cases, but inconsistently, while manual testing is accurate in all scenarios, ensuring a more rigorous verification aligned with good practice.

Both approaches maintain consistency in the names of test classes and methods, facilitating the identification of tested functionalities. For example, in the BankAccount tests, both manual and GPT-generated tests use names such as BankAccountTest.  

Both have clear codes in terms of structure, but the manual tests follow more organized patterns, with better separation and categorization of cases. In contrast, GPT tests may include redundancies or group tests in a confusing way, making it difficult to understand and detect errors. Thus, although both approaches are structured, manual tests stand out for their greater clarity and organization.

Both approaches handle exceptions correctly, but manual tests are more accurate by validating their occurrence and the exact message using assertEquals, ensuring strict verification when the message is crucial. In contrast, while checking the exception, GPT-generated tests do not always compare the message accurately. When they do, they use more steps, such as intermediate variables and assertTrue with contains, which provides flexibility but reduces rigor and can introduce unnecessary complexity.

Both approaches document tests well, with comments explaining their purpose and validations. However, some tests generated by GPT lack comments, although this appears to be an isolated error, as the model usually generates adequate documentation. Despite these occasional omissions, most of the comments in the automatically generated tests are accurate and fulfill their explanatory function.

\begin{figure} 
    \centering
    \resizebox{0.85\textwidth}{!}{ 
    \begin{tikzpicture}
    \begin{axis}[
        xbar, xmin=0, xmax=3.6, 
        width=12cm, height=6cm, enlarge y limits=0.2,
        symbolic y coords={Documentation, Coh. in names, Code structure, Numeric print, Exception handling, Boundary value, Equiv. partition},
        ytick=data,
        xtick={0.5,2,3.5}, 
        xticklabels={Insufficient, Adequate, Good},
        x tick style={draw=none}, 
        axis x line=bottom, 
        legend pos=outer north east, 
        bar width=5pt 
    ]
    \addplot[fill=white] coordinates {(3.5,Documentation) (3.5,Coh. in names) (2,Code structure) (2,Numeric print) (2,Exception handling) (2,Boundary value) (3.5,Equiv. partition)};
    \addplot[fill=black] coordinates {(3.5,Documentation) (3.5,Coh. in names) (3.5,Code structure) (3.5,Numeric print) (3.5,Exception handling) (3.5,Boundary value) (3.5,Equiv. partition)};

    \legend{LLM,Human}
    \end{axis}
    \end{tikzpicture}
    }
    \caption{Comparison of qualitative manual analysis}
    \label{fig:desempeno_llm_humano}
\end{figure}
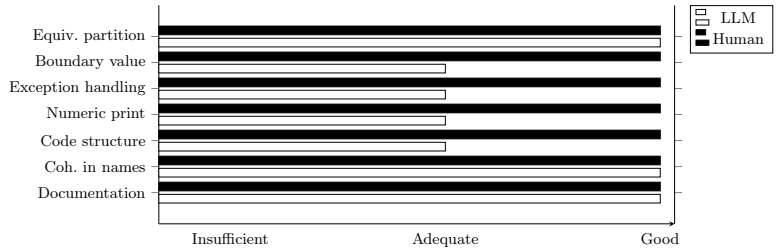

\section{Threads to Validity}
\label{section:threats}

The metrics used to evaluate the generated sets include code coverage, mutation score, and bad smells detection, all of which have been used in previous research, supporting their validity. However, a manual qualitative analysis, not used in similar studies, is incorporated. This approach allows the identification of aspects that automated metrics might miss, but it introduces a possible subjective bias inherent in human interpretation. Although this is minimized, it remains a limitation of qualitative analysis.

While this combination of approaches allows for a more comprehensive analysis of test quality, certain aspects of test effectiveness may not have been fully captured. Future studies could explore additional metrics or more specific criteria to strengthen the validity of the results.

The data obtained from the metrics were analyzed manually using tables and statistics, allowing for a structured comparison between the different approaches evaluated. No inconsistencies or significant variations in the metrics used were identified, suggesting stability in the results.

This research was based on 10 scenarios, which limits the generalization of the results. The choice of a reduced set allowed for a more detailed manual analysis, but implies that the findings may not generalize to larger test sets or other domains.

Furthermore, the results are limited to the Java language and the JUnit testing framework, so their extrapolation to other languages and tools is not guaranteed. Languages with different paradigms, such as Python with pytest or JavaScript with Jest, might require methodological adjustments.

The evaluated scenarios focused on equivalence partitions and boundary values without addressing other strategies such as structural, functional, error-based, or combinatorial testing. This could affect the applicability of the findings in contexts where a broader coverage of testing techniques is required. Although the LLM used showed no difficulties with code structure in these scenarios, no tests have been conducted in other environments to assess its performance on different software architectures or types of tests.

\section{Conclusions and Future Work}
\label{section:conclusions}

This paper presents an initial investigation into the use of general LLMs for automated unit test generation using the GPT-4 API. It evaluates its ability to generate quality test cases by identifying equivalence partitions and boundary values. The results show that, in one case, the LLM does not generate executable tests without human intervention. At the same time, in three scenarios, errors arise from inaccuracies in the assertions with float and double types. Generally, coverage, quality, and bad smell metrics are similar in manual and automated tests. However, manual qualitative analysis is key to detecting aspects that automated analysis might miss.

Based on the observed results, it is not considered that an LLM can write unit tests fully autonomously; it still requires human supervision. However, a tester with solid technical knowledge can guide the LLM with precise indications to generate effective tests. This suggests that, in the future, testers could focus on designing prompts and scripts, delegating the generation of test code to artificial intelligence, thus reducing the need for advanced technical skills. 

A line of future work could explore LLMs' capability to analyze and correct their errors and assess their ability to self-adjust. Another area of future research is to examine how the language of the prompts influences the quality of the tests, considering that some languages have a higher representation in the training data. Furthermore, it is proposed to explore the impact of sampling temperature (0.3 in this study) on the accuracy and creativity of the generated tests.

\bibliographystyle{splncs04}
\bibliography{bibliography}

\end{document}